\documentclass{frontiersFPHY} 

\usepackage{amsmath}
\usepackage{bbm}
\usepackage{hyperref}
\usepackage{amssymb}
\newcommand{\1}{\mathbbm{1}}
\newcommand{\N}{\mathcal{N}}
\newcommand{\ket}[1]{|{#1}\rangle}
\newcommand{\dyad}[2]{|{#1}\rangle \langle #2|}

\usepackage{url,hyperref,lineno,microtype}
\usepackage[onehalfspacing]{setspace}

\def\cotan{\qopname\relax o{cotan}}

\def\keyFont{\fontsize{8}{11}\helveticabold }
\def\firstAuthorLast{Coulamy et al.} 
\def\Authors{$^{1}$Ivan B. Coulamy, $^1$Alan C. Santos, $^{2}$Itay Hen and $^{1,3,*}$Marcelo S. Sarandy}
\def\Address{$^{1}$Instituto de F\'{i}sica, Universidade Federal Fluminense, Av. Gal. Milton Tavares de Souza s/n, Gragoat\'{a}, 24210-346 Niter\'{o}i, Rio de Janeiro, Brazil \\
$^{2}$Information Sciences Institute, University of Southern California, Marina del Rey, California 90292, USA \\
$^{3}$Center for Quantum Information Science \& Technology and Ming Hsieh Department of Electrical Engineering, University of Southern California, Los Angeles, California 90089, USA}
\def\corrAuthor{Marcelo S. Sarandy}
\def\corrAddress{Instituto de F\'{i}sica, Universidade Federal Fluminense, Av. Gal. Milton Tavares de Souza s/n, Gragoat\'{a}, 24210-346 Niter\'{o}i, Rio de Janeiro, Brazil}
\def\corrEmail{msarandy@if.uff.br}

\begin{document}
\onecolumn
\firstpage{1}

\title{Energetic cost of superadiabatic quantum computation} 

\author[\firstAuthorLast ]{\Authors} 
\address{} 
\correspondance{} 

\extraAuth{}

\maketitle


\begin{abstract}
We discuss the energetic cost of superadiabatic models of quantum computation. Specifically, 
we investigate the energy-time complementarity in general transitionless controlled evolutions and 
in shortcuts to the adiabatic quantum search over an unstructured list. We show that the additional 
energy resources required by superadiabaticity for arbitrary controlled evolutions can be minimized by 
using probabilistic dynamics, so that the optimal success probability is fixed by the choice of the 
evolution time. In the case of analog quantum search, we show that the superadiabatic 
approach induces a non-oracular counter-diabatic Hamiltonian, with the same energy-time complexity 
as equivalent adiabatic implementations. 
\tiny
 \keyFont{ \section{Keywords:} Quantum Computing, Quantum Information, Shortcuts to Adiabaticity, Superadiabaticity, Quantum Gates, Quantum Search} 
\end{abstract}

\section{Introduction}
Shortcuts to adiabatic passage~\cite{Demirplak:03,Demirplak:05,Berry:09,Torrontegui:13} provide a 
remarkable mechanism for speeding up quantum tasks, which can be achieved through the use of 
a counter-diabatic assistant driving. These techniques have been introduced to mimic the transitionless 
adiabatic dynamics, but with the usual constraint on the adiabatic runtime lifted. Transitionless 
quantum driving has been applied to a number of quantum information protocols, such as population 
transfer~\cite{Chen:14-PRA,Lu:14-LP} and entanglement 
generation~\cite{Lu:14-PRA,Chen:14-LPL,Chen:15-PRA,Chen:15-SciRep}. In the context of 
many-body systems, realizable settings have been investigated for assisted evolutions in 
quantum critical phenomena~\cite{Campo:12,Campo:13,Saberi:14}.  More recently, counter-diabatic 
approaches have been proposed for fast implementation of individual unitaries in quantum circuits, 
leading to universal {\it superadiabatic} schemes of quantum computing (QC) via local Hamiltonians~\cite{SciRep:15,Santos:16}. Such methods may be potentially relevant to 
accelerating the implementation of $n$-qubit controlled gates in digitized proposals of adiabatic quantum 
computing (see, e.g., Refs.~\cite{Itay:15, Kieferova:14, Martinis:14, Barends:15}).

The superadiabatic speedup is intrinsically connected with an increase of the energy resources 
demanded by the quantum computer~\cite{SciRep:15,Santos:16}, which in turn implies a 
rather versatile computational cost that is controlled by the energetic capacity available to the physical 
apparatus. Here we show that this energy-time complementarity can be exploited in 
quantum information processing. First, we consider controlled evolutions (CE) as a mechanism 
to implement superadiabatic universal QC~\cite{SciRep:15} which generalizes the original adiabatic 
approach introduced in Ref.~\cite{Itay:15}. We then show that, within the superadiabatic scenario, the 
energetic cost can be minimized by replacing the deterministic realization of quantum gates for probabilistic implementations based on a probability distribution of a binary random variable described by an angle parameter. 
By doing so, the energy expense can be minimized by adjusting the probability distribution, provided the choice of the 
evolution time of the computational process. Second, we analyze the effects of the energy-time complementarity in 
analog quantum search~\cite{grover}, where the oracular approach designed by the local adiabatic Grover algorithm
is known to be optimal~\cite{Dam:01,Roland:02}. In this case, we show that the superadiabatic approach 
naturally requires an unphysical non-oracular counter-diabatic Hamiltonian, 
with the energy-time complexity equivalent to non-oracular adiabatic implementations. 

The paper is organized as follows. In Section 2, we describe the adiabatic implementation of 
quantum gates via CE and several adiabatic quantum search approaches. We then provide their 
superadiabatic versions and introduce the metric for energetic cost used in our work. In Section 3, we 
investigate the energy complexity of the superadiabatic realizations of both quantum gates via CE and analog 
quantum search. In particular, we consider the properties of the probabilistic model of QC through CE 
and the consequences of the energy-time complementarity for the search problem. 
In Section 4, we present our conclusions and future perspectives. 

\section{Methods}

Our aim in this Section is to discuss adiabatic implementations of QC, their superadiabatic generalizations, 
and the energetic cost measure adopted in this work.   

\subsection{Quantum gates by adiabatic controlled evolutions} \label{AdiaGate}

Let us begin by using adiabatic CE~\cite{Itay:15} to implement $n$-controlled gates~\cite{SciRep:15}. 
To this end, we will consider the adiabatic evolution of a composite system $\mathcal{T} \mathcal{A}$ associated with 
a Hilbert space ${\mathcal H}_{\mathcal T} \otimes {\mathcal H}_{\mathcal A}$, where 
$\mathcal{T}$ denotes a target subsystem containing $n+1$ qubits and $\mathcal{A}$ denotes 
an auxiliary subsystem containing a single qubit. We will use the first $n$ qubits of $\mathcal{T}$ as 
the control register of the $n$-controlled gate, while the last qubit will play the role of 
its target register. Then, a rotation of the target qubit of an angle $\phi $ around a direction $\hat{n}$ 
in the Bloch sphere will be performed when the state of the control register is 
$\left\vert 11\cdots 1\right\rangle$. We will adopt here the decimal representation 
$\left\vert 11\cdots 1\right\rangle \equiv \left\vert N-1\right\rangle$, with $N=2^n$. 
An $n$-controlled rotation over a single qubit can be adibatically implemented by preparing the auxiliary
qubit in the initial state $\left\vert 0\right\rangle$, with the
adiabatic Hamiltonian given by~{\cite{SciRep:15}}%
\begin{equation}
H\left( s\right) =\left[ \1-P_{N-1,n_{-}}\right] \otimes H_{0}\left( s\right)
+P_{N-1,n_{-}}\otimes H_{\phi }\left( s\right) \text{ \ ,}  \label{HAN}
\end{equation}%
where $P_{k,{n}_{\pm }}=\left\vert k\right\rangle \left\langle
k\right\vert \otimes \left\vert \hat{n}_{\pm }\right\rangle \left\langle 
\hat{n}_{\pm }\right\vert $ is the set of all orthogonal projectors on the
subspace $\mathcal{T}$ and $\left\vert \hat{n}_{\pm }\right\rangle
\left\langle \hat{n}_{\pm }\right\vert =1/2\left( \1 \pm \hat{n}\cdot \vec{%
\sigma}\right) $, with $\vec{\sigma}=\left( \sigma _{x},\sigma _{y},\sigma
_{z}\right) $. The Hamiltonians $H_{0}\left( s\right) $ and $H_{\phi
}\left( s\right) $ are given by%
\begin{equation}
H_{\xi }\left( s\right) =-\hbar \omega \left\{ \sigma _{z} \cos \theta \left( s\right)
+\sin \theta \left( s\right) [ \sigma _{x}\cos \xi +\sigma
_{y}\sin \xi ] \right\}  \text{ \ ,}  \label{Hxi}
\end{equation}%
where $\theta \left( s\right) =\theta _{0}s$, $\theta_0$ is a constant angle, $\xi = \left\{ 0,\phi \right\}$, and $s$ denotes the normalized time $s=t/\tau$, with 
$\tau$ the total evolution time. The system is prepared in the initial state $\left\vert \Psi
\left( 0\right) \right\rangle =\left\vert \psi _{n}\right\rangle \otimes \left\vert
0\right\rangle $, where 
\begin{equation}\left\vert \psi _{n}\right\rangle
=\sum_{m=0}^{N-1}\sum_{\epsilon=\pm }\gamma
_{m,\epsilon }\left\vert m,\hat{n}_{\epsilon }\right\rangle  \text{ \ .} \label{InitialState}
\end{equation} 
Then, by adiabatic evolution, the system will evolve to the final state
$\left\vert \Psi \left( 1\right) \right\rangle$ given by%
\begin{equation}
\left\vert \Psi \left( 1\right) \right\rangle =\left[ \left(
\1 - P_{N-1,\hat{n}_{-} }\right) \left\vert \psi _{n}\right\rangle \right] \otimes
\vert E_{0}^{0}\left( 1\right) \rangle +P_{N-1,\hat{n}_{-} }\left\vert
\psi _{n}\right\rangle \otimes \vert E_{\phi }^{0}\left( 1\right)
 \rangle \text{ \ ,}   \label{PSI2}
\end{equation}%
where $\vert E_{\xi }^{0}\left( 1\right) \rangle =\cos \left(
\theta _{0}/2\right) \left\vert 0\right\rangle +e^{i\xi }\sin \left( \theta
_{0}/2\right) \left\vert 1\right\rangle $ is the ground state of $H_{\xi }\left( 1\right)$. Then, equivalently, 
we can write
\begin{equation}
|\Psi \left( 1\right) \rangle =\cos \left( \theta _{0}/2\right) \left\vert \psi _{n}\right\rangle \otimes \vert 0 \rangle +\sin \left( \theta _{0}/2\right) \vert \psi _{n}^{\text{rot}}\rangle \otimes
\left\vert 1\right\rangle  \text{ \ ,}  \label{PSI1}
\end{equation}%
with 
\begin{equation}
|\psi _{n}^{\text{rot}}\rangle =\sum_{k=0}^{N-2}\sum_{\epsilon = \pm }\gamma _{k, \epsilon
}\left\vert k , \hat{n}_{\epsilon} \right\rangle +\left\vert N-1\right\rangle \otimes [ \gamma _{N-1,{+}}\left\vert \hat{n}_{+}\right\rangle +e^{i\phi
}\gamma _{N-1,{-} }\left\vert \hat{n}_{-}\right\rangle ]  \text{ \ .} \label{rotatedState}
\end{equation}%

The rotated state $\vert \psi _{n}^{\text{rot}} \rangle$ is the target of the $n$-controlled gate. However, note 
that $\left\vert \Psi \left( 1\right) \right\rangle $ in Eq.~(\ref{PSI1}) is
an entangled state. Thus a measurement must be performed on the auxiliary system,
where the action of the gate will be considered successful if ${\mathcal{A}}$ is measured in the state $|1\rangle$, which occurs with probability $\sin ^{2}\left( \theta_{0}/2\right) $. On the other hand, if the outcome of a measurement on $\mathcal{A}$ yields $\left\vert 0 \right\rangle$, the adiabatic evolution should be restarted through the 
Hamailtonian in Eq.~(\ref{HAN}), as the state of the system is projected onto the initial state $\left\vert \Psi \left( 0\right)
\right\rangle $. Naturally, by choosing $\theta _{0}=\pi$, we deterministically ensure the success of the computation. 
However, as we will show, deterministic evolutions may demand more energy resources than probabilistic processes 
when transitionless drivings are considered. 
In particular, observe also that the scheme presented here allows for the implementation of arbitrary $n$-controlled gates, which lead to versatile sets of universal gates, e.g., single qubit rotations and controlled-NOT operations~\cite{nielsen}. \\

\subsection{Adiabatic quantum search} 

Instead of adiabatic implementations of quantum circuits, we can also consider the original approach of 
adiabatic QC~\cite{Farhi:Science}, where a single annealing process is performed using energy penalties 
attributed to quantum states that violate the solutions of an optimization problem. Here we employ this 
method to analyze three possible adiabatic implementations of quantum search over an unstructured list. 
An adiabatic QC approach for the quantum search through Grover's algorithm~\cite{grover} 
was first proposed in Ref.˜\cite{farhi1} and improved by using local adiabaticity~\cite{Dam:01,Roland:02}, where 
the adiabatic evolution is required for each local time interval, instead of being globally applied as in the 
original proposal. In both cases, the search for a marked element in an unstructured list 
of $N=2^n$ elements (labeled by $n$ qubits) can be achieved by employing a Hamiltonian of the form
\begin{equation} \label{eq:grover-hamiltoniano}
H_0(s) = f(s)(\1 - | + \rangle \langle +|) +g(s)(\1-|m \rangle \langle m|), 
\end{equation}
where $|m\rangle$ is the marked state, $s$ is the normalized time ($0\le s \le 1$), $| {+}\rangle = 1/\sqrt{N}\sum_{i=0}^{N-1}|{i}\rangle$, and $f(0)=g(1)=1$ and $f(1)=g(0)=0$.  The eigenspectrum of this Hamiltonian can be exactly derived (see, e.g., Ref.~\cite{das,Orus:04}). 
In particular, the two lowest eigenstates can be written as
\begin{equation} \label{eq:eivec}
|{E_\pm(s)}\rangle = \N_{\pm}(s) \left[ |{m}\rangle + b_\pm(s)\ket \phi \right],
\end{equation}
where the normalization constant is ${\mathcal N}_{\pm}(s) =1/{\sqrt{1+(N-1)b_\pm(s)^2}}$, $\ket \phi=\sum_{i \neq m} |{i}\rangle$, and 
\begin{equation}
b_\pm(s) =1-\frac{E_\pm(s)}{f(s)\overline N},
\end{equation} 
with $\overline N = 1-1/N$, and the corresponding energies $E_\pm(s)$ given by
\begin{equation} \label{Ener-lowest-G}
 E_\pm(s) =\frac{f(s)+g(s) \pm \sqrt{[f(s)+g(s)]^2-4 f(s)g(s)\overline N}}{2}.
 \end{equation}
The other higher-energy eigenstates form an 
$(N-2)$-fold degenerate eigenspace, whose energy is given by 
\begin{equation} \label{Ener-deg-G}
E_{\text{deg}}=\left[f(s)+g(s)\right].
\end{equation}
In order to explicitly provide the eigenstates $|E_{\text{deg}}^k\rangle$ ($k=1,\cdots,N-2$) associated with the eigenenergy $E_{\text{deg}}$, we write 
\begin{equation} \label{E-deg-Grover}
|E_{\text{deg}}^k\rangle = \sum_{n=0}^{N-1} c^k_n |n\rangle .
\end{equation}
Then, from the eigenvalue equation for $H_0(s)$, it directly follows that the set $\{c^k_n\}$ is just required to 
satisfy the constraints $\sum_{n=0}^{N-1} c^k_n = 0$ and $c^k_m=0$. As a consequence, 
the states $|E_{\text{deg}}^k\rangle$ can be suitably chosen as {\it{time-independent}} vectors.

By imposing a local adiabatic evolution~\cite{Dam:01,Roland:02}, i.e. by requiring adiabaticity at each infinitesimal time interval, the runtime is minimized for the path (see also, e.g.,  Ref.~\cite{Kieferova:14})
\begin{equation} \label{LAE-Grover}
f(s)=1-g(s), \,\,\,\,\,\,\,\,\, g(s) = \frac{\sqrt{N-1} - \tan\left[\arctan\left(\sqrt{N-1}\right)\left(1-2s\right)\right]}{2\sqrt{N-1}}.
\end{equation}
This results in a quadratic speedup over the classical search, i.e., we obtain the 
time complexity $O(\sqrt{N})$ expected by the Grover quantum search~\cite{Dam:01,Roland:02}. 

It is possible to reduce the time complexity of the Grover quantum search by transferring the algorithmic cost to other physical resources. The second implementation of the adiabatic Grover search considered here has been 
introduced in Ref.~\cite{das,Wen:08}. It is also based on the Hamiltonian in Eq.~(\ref{eq:grover-hamiltoniano}) 
to perform the evolution, but requiring that the functions $f(s)$ and $g(s)$ satisfy
\begin{eqnarray} 
f(s) &=& 1-s + \sqrt{N}(1-s)s , \label{eq:f-def}
 \\
g(s) &=& s + \sqrt{N}(1-s)s . \label{eq:g-def}
\end{eqnarray}
This implementation achieves the solution at constant time complexity $O(1)$. As is apparent from 
Eqs.~(\ref{eq:f-def}) and (\ref{eq:g-def}), the original time resource has been transferred to the coupling strengths $f(s)$ and $g(s)$ and as discussed in detail in the next Section, 
will be reflected in the energy scaling required by the system. 

The two previous versions of the adiabatic Grover's algorithm are based on oracular Hamiltonians, which we take here to be operators able to recognize the correct answer of a problem~\cite{nielsen}. This 
is indeed the case if one chooses a Hamiltonian composed of an operator $O_m$ in the form
$O_m = \1 - |m\rangle \langle m |$.
The action of $O_m$ in the computational basis $\{|i\rangle\}$ is
\begin{equation}
O_m \ket i = (\1 - |m\rangle \langle m |) \ket{i} = \begin{cases}
0
& (i = m) ,
\\
 \ket{i}
& (i \neq m),
\end{cases} \label{eq:oracular}
\end{equation}
so that this operator recognizes the marked state, providing no hint about its identity if acting upon any other state.
Adiabatic versions of the quantum search have also been proposed via non-oracular Hamiltonians. Our 
third implementation of Grover's algorithm is based on the 
non-linear non-oracular (NLNO) Hamiltonian proposed in Ref.~\cite{wen}. In this work, the time-dependent 
Hamiltonian in Eq.~\eqref{eq:grover-hamiltoniano} is replaced for
\begin{equation} 
H_0(s) = f(s)(\1 - | + \rangle \langle +|) +g(s)(\1-|m \rangle \langle m|) + h(s)( | + \rangle \langle m|+ |m \rangle \langle +|), 
\label{NOH}
\end{equation}
where $h(0)=h(1)=0$.
The Hamiltonian in Eq.~(\ref{NOH}) contains an operator \hbox{$\overline{O}_m = |+ \rangle \langle m | + |m\rangle \langle + |$}. The action of $\overline{O}_m$ in the computational 
basis $\{|i\rangle\}$ is
\begin{equation}  \label{NO-eq}
\overline{O}_m \ket i = (\dyad{+}{m} + \dyad{m}{+}) \ket{i} = \begin{cases}
\frac{1}{\sqrt{N}}\ket{m}+\ket {+}
& (i = m), 
\\
\frac{1}{\sqrt{N}} \ket{m}
& (i \neq m) .
\end{cases} 
\end{equation}
Observe that Eq.~\eqref{NO-eq} implies that 
$\overline{O}_m$ cannot {\it exactly} 
recognize a marked element, even though it could  {\it effectively} recover the marked state for $N\gg 1$ 
with a single operation over the uniform superposition provided by the state $|+\rangle$.  
Naturally, the non-oracular form of the Hamiltonian involves all the individual 
computational states, requiring therefore much more than the capacity of the Hamiltonian to recognize the marked element.   
This is an obviously artificial approach, whose discussion here is kept just for comparison with the superadiabatic 
scenario. Assuming a restricted feasibility of such a Hamiltonian, we proceed by looking at its eigenspectrum. The 
ground and first excited states have the same structure as in Eq.~\eqref{eq:eivec}, with
\begin{equation}
b_\pm(s) = \frac{\overline{N} f(s)+\frac{2 h(s)}{\sqrt{N} } - E_\pm(s)}{\overline{N}\left[ f(s)-h(s) \sqrt N\right]}.
\end{equation}
The two lowest energy levels are given by
\begin{eqnarray}
E_\pm(s) &=& \frac{1}{2} \left\{ f(s)+g(s) +\frac{2h(s)}{\sqrt{N}} \right. \nonumber \\
&& \left. \pm \sqrt{\left[f(s)+g(s)\right]^2 - 4 f(s)g(s) \overline N +4h^2(s)-\frac{4h(s)}{\sqrt{N}}\left[f(s)+g(s)\right]} \, \right\}. \label{NOH-epm}
\end{eqnarray}
As before, the higher-energy states form an $(N-2)$-fold degenerate subspace, with energy given by $f(s)+g(s)$. 
As shown in Ref.~\cite{wen}, this formulation also shows  
constant time complexity $O(1)$, which can be obtained by choosing a suitable interpolation, such as
\begin{equation} \label{NOH-inter}
f(s)=1-s, \,\,\,\,\,\,\,\, g(s)=s, \,\,\,\,\,\,\,\, h(s)=s(1-s).
\end{equation}

\subsection{Speeding up adiabaticity through superadiabatic evolutions}

The performance of adiabatic QC is dictated by a long total evolution time compared to the inverse of a power of the 
energy gap~\cite{Messiah:book,Teufel:03,Jansen:07,Sarandy:04}. However, the adiabatic evolution can be sped up through shortcuts to adiabaticity via counter-diabatic Hamiltonians~\cite{Demirplak:03,Demirplak:05,Berry:09}. The fundamental idea underlying these shortcuts to adiabaticity is to add a new contribution $H_{\text{CD}}\left( t\right) $, called \textit{counter-diabatic Hamiltonian}, to the original adiabatic Hamiltonian $H\left( t\right)$. 
This term is constructed such that it allows the mimicking of the adiabatic evolution, however without any constraint on 
the total time of evolution. The total composite Hamiltonian is
\begin{equation}
H_{\text{SA}}\left( t\right) =H\left( t\right) +H_{\text{CD}}\left( t\right)  \text{ \ ,}
\label{SH}
\end{equation}%
which is called \textit{superadiabatic Hamiltonian}. In particular, it is possible to show that the counter-diabatic 
term reads~\cite{Berry:09}
\begin{equation}
H_{\text{CD}}\left( t\right) =i\hbar \sum\nolimits_{n}\left\vert \dot{n}\left(
t\right) \right\rangle \left\langle n\left( t\right) \right\vert
+\left\langle \dot{n}\left( t\right) |n\left( t\right) \right\rangle
\left\vert n\left( t\right) \right\rangle \left\langle n\left( t\right)\right\vert  \text{ \ ,}  \label{CDH}
\end{equation}%
where $\left\vert n\left( t\right) \right\rangle $ is the eigenstate of $%
H\left( t\right) $ associated to the energy $E_{n}\left( t\right) $. 
The goal of the counter-diabatic
term $H_{\text{CD}}\left( t\right) $ in the Hamiltonian $H_{\text{SA}}\left( t\right) $\
is exactly to eliminate the diabatic contributions of $H(t)$.
Thus, if the system is initially prepared in the ground state of $%
H\left( 0\right) $, then the system will deterministically evolve to the instantaneous ground 
state of the Hamiltonian $H\left( t\right)$ with no constraints over the evolution time.
Note that, in general, one would need to be able to explicitly calculate all the eigenstates of $H\left( t\right)$ to derive 
a shortcut to adiabaticity using the counter-diabatic driving. However, this may not be a hard 
requirement in the case of superadiabatic versions of circuit implementations, where one-qubit rotations and 
two-qubit entangling gates are enough to achieve QC universality~\cite{nielsen}. 
In particular, as we shall see for this case,  
$H_{\text{CD}}\left( t\right)$ can be realized through a simple time-independent operator.

\subsection{Energetic cost of quantum evolutions}

To quantify the expense of energy in a quantum evolution driven by a Hamiltonian $H(t)$, 
we adopt as the cost measure the average norm of $H(t)$ computed for a total time of evolution $\tau$. 
This yields~\cite{SciRep:15,Santos:16,Kieferova:14,Zheng:15}
\begin{equation}
\Sigma \left( \tau \right) =\frac{1}{\tau }\int_{0}^{\tau }\left\Vert
H\left( t\right) \right\Vert dt=\int_{0}^{1}\left\Vert H\left( s\right)
\right\Vert ds \text{ \ ,}  \label{cost.1}
\end{equation}%
where $s=t/\tau$ is the parametrized time and the norm here is defined by the Frobenius norm (Hilbert-Schmidt norm) $\left\Vert
A\right\Vert =\sqrt{\text{Tr}\left[ A^{\dagger }A\right] }$. Naturally, other norms can be adopted as, for instance,  
the spectral norm 
$\left\Vert A\right\Vert_{2} =\sqrt{\lambda_{\max} \left[ A^{\dagger }A\right]}$, where $\lambda_{\max}$ denotes the maximum eigenvalue of $\left[ A^{\dagger }A\right] $. For the Hamiltonians investigated in this work, these norms 
will imply into a cost simply related by a constant $D^{1/2}$, with $D$ denoting the dimension of corresponding the Hilbert space. 
The Frobenius norm as well as arbitrary superadiabatic 
evolutions with total evolution time $\tau$, the energetic cost can be written as 
\begin{equation}
\Sigma_{SA} \left( \tau \right) =\frac{1}{\tau }\int_{0}^{\tau }\sqrt{\text{Tr}%
\left[ {H}_{SA}^{2}\left( t\right) \right] }dt  =\frac{1}{\tau }\int_{0}^{\tau }\sqrt{\text{Tr}%
\left[ {H}^{2}\left( t\right) + {H}_{CD}^{2}\left( t\right) \right] }dt , \label{cost1.2}
\end{equation}
where we have used that ${\textrm{Tr}} \left( \left\{ H(t), \frac{}{} \hspace{-0.05cm}H_{CD}(t) \right\} \right) = 0$~\cite{SciRep:15}. 
This explicitly shows that a superadiabatic evolution has an energetic cost larger than its corresponding 
adiabatic evolution. By evaluating the trace in Eq.~(\ref{cost1.2}), we obtain
\begin{equation}
\Sigma _{\text{SA}}\left( \tau \right) = \int_{0}^{1}\sqrt{\sum_{m}%
\left[ E_{m}^{2}\left( s\right) +\hbar^2 \frac{\mu _{m}\left( s\right) }{\tau^{2}}\right] }ds \text{ \ ,}  \label{cost.5}
\end{equation}%
where $\mu _{m}\left( s\right) =\left\langle \partial _{s}m\left( s\right)
|\partial _{s}m\left( s\right) \right\rangle -\left\vert \left\langle
m\left( s\right) |\partial _{s}m\left( s\right) \right\rangle \right\vert
^{2}$ and $\{E_{m}\left( s\right) \}$ is the energy spectrum of the adiabatic Hamiltonian $H(t)$, 
with $\{|m(s)\rangle\}$ denoting its eigenbasis. Notice that the adiabatic limit is recovered 
when taking $\tau \rightarrow \infty$.
Thus, the speedup obtained by the superadiabatic dynamics is limited by the energetic cost 
of the evolution. Indeed, this energy-time complementarity can be formally discussed through the 
quantum speed limit~\cite{Deffner:13}, which suggests that the superadiabatic evolution time 
is compatible with arbitrarily short time intervals (implying into corresponding arbitrarily large energies)~\cite{SciRep:15}, 
while the adiabatic evolution time obeys the 
lower bound $\tau _{\text{Ad}}\propto 1/\omega^{n}$, with $\omega$ associated with the energy gap and $%
n\in \mathbb{N}^{+}$ \cite{Messiah:book,Teufel:03,Jansen:07,Sarandy:04}. 

\section{Results}

We now consider the performance of adiabatic and superadiabatic quantum computation, focusing on their 
time-energy complexity. This will be investigated both for the universal model of superadiabatic 
controlled gates and for the superadiabatic implementations of the Grover search. 

\subsection{Quantum gates by superadiabatic controlled evolutions} \label{CSEandUQC}

Let us begin by discussing the superadiabatic model of universal QC via CE implemented by shortcuts to adiabaticity~\cite{SciRep:15}. To this end, let us first write 
the complete set of eigenstates of $H\left( t\right)$ as~\cite{SciRep:15}
\begin{eqnarray}
|E_{0m}^{\epsilon k}\left( s\right) \rangle  &=&\left\vert m,\hat{n}%
_{\epsilon }\right\rangle \otimes |E_{0}^{k}\left( s\right) \rangle \text{ \
,}  \label{Ev1} \\
|E_{0\,(N-1)}^{+k}\left( s\right) \rangle  &=&\left\vert N-1,\hat{n}%
_{+}\right\rangle \otimes |E_{0}^{k}\left( s\right) \rangle \text{ \ ,}
\label{Ev2} \\
|E_{\phi \,(N-1)}^{-k}\left( s\right) \rangle  &=&\left\vert N-1,\hat{n}%
_{-}\right\rangle \otimes |E_{\phi }^{k}\left( s\right) \rangle \text{ \ ,}
\label{Ev3}
\end{eqnarray}%
where $m \in \{0,\cdots,N-2\}$, $\epsilon,k \in \{\pm\}$, and 
\begin{eqnarray}
|E_{\xi }^{+}\left( s\right) \rangle  &=&-\sin \frac{\theta _{0}s}{2}%
\left\vert 0\right\rangle +e^{i\xi }\cos \frac{\theta _{0}s}{2}\left\vert
1\right\rangle \text{ \ ,}  \label{cqa.2.5a} \\
|E_{\xi }^{-}\left( s\right) \rangle  &=&\cos \frac{\theta _{0}s}{2}%
\left\vert 0\right\rangle +e^{i\xi }\sin \frac{\theta _{0}s}{2}\left\vert
1\right\rangle \text{ \ ,}  \label{cqa.2.5b}
\end{eqnarray}%
with $\xi \in \{0,\phi\}$ and $\{|E_{\xi }^{\pm}\left( s\right) \rangle\}$ denoting the set of eigenstates of 
each adiabatic Hamiltonian $H_{\xi }\left( s\right)$, as provided by Eq.~(\ref{Hxi}).
Thus, by using the Eq.~(\ref{SH}), we can show that the superadiabatic Hamiltonian is given by~\cite{SciRep:15}
\begin{equation}
H_{\text{SA}}\left( s\right) =\left[ 1-P_{N-1,\hat{n}_{-} }\right] \otimes
H_{0}^{\text{SA}}\left( s\right) +P_{N-1,\hat{n}_{-} }\otimes H_{\phi }^{\text{SA}}\left(
s\right)  \text{ \ ,}  \label{HSAN}
\end{equation}%
where each term $H_{\xi }^{\text{SA}}\left( s\right)$ corresponds to the superadiabatic
Hamiltonian associated with the adiabatic Hamiltonian $H_{\xi }\left( s\right)$, i.e.  
$H_{\xi }^{\text{SA}}\left( s\right) =H_{\xi }\left( s\right)
+H_{\xi }^{\text{CD}} $, with 
\begin{equation}
H_{\xi }^{\text{CD}} =\hbar \frac{\theta _{0}}{2\tau }\left( \sigma
_{y}\cos \xi -\sigma _{x}\sin \xi \right)  \label{CDIndTempo}
\end{equation}%
being the (time-independent) counter-diabatic contribution to achieve the evolution at total time $\tau$~{\cite{SciRep:15}}.

\subsection{Energy-time complementarity in the CE model of quantum gates}

Let us now consider Eq.~(\ref{cost.5}) to investigate the time-energy complementarity relationship in both  adiabatic and superadiabatic CE models of universal quantum gates. 
To this end, we need the set of eigenvalues and eigenstates of the adiabatic Hamiltonian in Eq.~(\ref{HAN}), which are given by Eqs. (\ref{Ev1})-(\ref{cqa.2.5b}). The spectrum of $H\left( s\right) $ has 
$\left( 2N\right) $-degenerate levels, with $%
\{|E_{0m}^{\epsilon +}\left( s\right) \rangle ,|E_{0\,(N-1)}^{++}\left(
s\right) \rangle ,|E_{\phi \,(N-1)}^{-+}\left( s\right) \rangle \}$ and $%
\{|E_{0m}^{\epsilon -}\left( s\right) \rangle ,|E_{0\,(N-1)}^{+-}\left(
s\right) \rangle ,|E_{\phi \,(N-1)}^{--}\left( s\right) \rangle \}$
associated with the levels $E^{+}=\hbar \omega $ and $E^{-}=-\hbar \omega $,
respectively. So, by using Eqs.~(\ref{Ev1})-(\ref{cqa.2.5b}), we can show that $\mu
_{l}^{m}\left( s\right) =\theta _{0}^{2}/4\tau ^{2}$. In addition, the 
energetic cost to implement any gate controlled by $n$ qubits is 
$\Sigma
_{\text{SA}}\left( \tau ,n\right) =2^{n/2}\Sigma _{\text{SA}}^{sing}\left( \tau \right) $~\cite{SciRep:15}, 
where $\Sigma _{\text{SA}}^{sing}$ 
 is the energetic cost to implement any single qubit unitary transformation, with
\begin{equation}
\Sigma _{\text{SA}}^{sing}\left( \omega \tau ,\theta _{0}\right) =2 \hbar \omega\sqrt{1+\frac{\theta
_{0}^{2}}{4\left( \omega \tau \right) ^{2}}} \text{ \ .}
\label{cost1.4}
\end{equation}%

\begin{figure}[h!]
\begin{center}
\includegraphics[width=12.5cm]{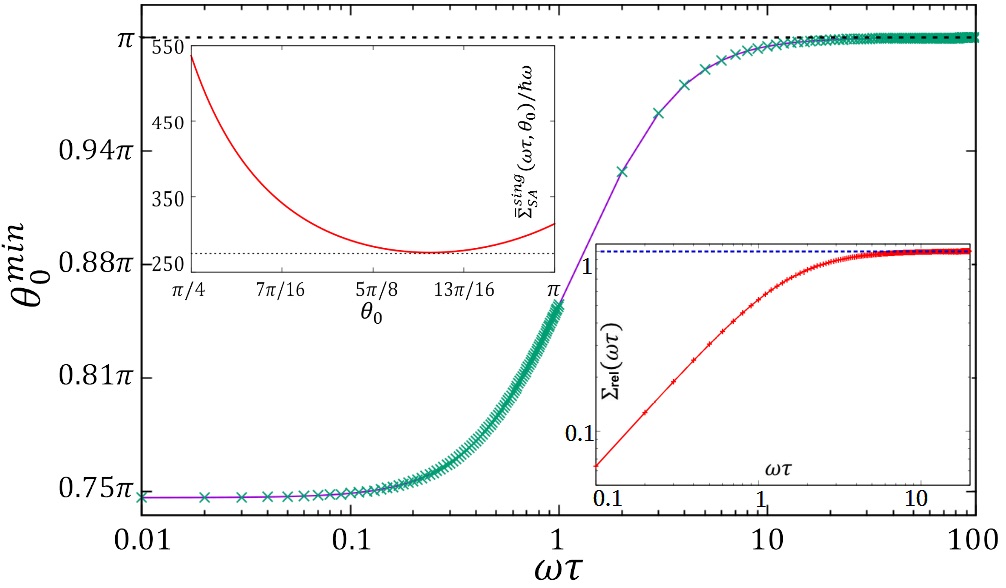}
\end{center}
 \textbf{\refstepcounter{figure}\label{fig1} Figure \arabic{figure}.}{ Optimal value $\theta _{0}^{\min }$ 
for the angle parameter $\theta_0$ as a function of $\omega \tau$, with $\omega \tau$ in logarithmic 
scale. The points are obtained from Eq.~(\ref{wtTheta}), with the curve denoting the numerical fit.
{\bf Upper inset}: Average energy in units of $\hbar \omega$ 
as a function of $\theta_0$ for $\omega \tau = 0.01$. The results are obtained from Eq.~(\ref{MediaCost}). 
{\bf Lower inset}: Fraction $\Sigma _{\text{rel}}\left( \omega \tau \right)$ of energy required by 
the optimized probabilistic model as a function of $\omega \tau$, with data in logarithmic scale. 
The points are obtained from Eq.~(\ref{sigma_rel}), with the curve denoting the numerical fit.
 } 
\end{figure}

A similar result can be obtained from the spectral norm, with energetic cost given by 
$\Sigma _{\text{SA}}^{sing}\left( \omega \tau ,\theta _{0}\right)\vert_{2} = (1/2) \Sigma _{\text{SA}}^{sing}\left( \omega \tau ,\theta _{0}\right)$, since the Hilbert space has dimension $D=4$ in this case. 
Note that the energetic cost is independent of the parameter $\theta_0$ in the adiabatic limit 
$\omega\tau \rightarrow \infty$. Therefore, the best computational adiabatic strategy is to set $\theta=\pi$, 
which deterministically ensures the implementation of the gate with probability one. On the other hand, 
probabilistic quantum computation can be energetically favored in the superadiabatic regime. 
Indeed, from Eq.~(\ref{PSI1}), we can see that, by setting  $0 < \theta_{0} < \pi $, the implementation of the 
quantum gate is achieved with a nonvanishing probability. Thus, we can investigate whether or not it is 
possible to find out a specific value of $\theta _{0}$ such that the energetic cost is better in average 
than the deterministic choice $\theta _{0}=\pi$. To address this point, let us define the quantity%
\begin{equation}
\langle N \rangle =\frac{1}{\sin ^{2}\left( \theta _{0}/2\right) } \text{ \ ,}
\label{MediaN}
\end{equation}%
which is the average number of evolutions for a successful computation. So, the average energetic cost to implement a probabilistic evolution is 
\begin{equation}
\bar{\Sigma}=\langle N \rangle \Sigma,
\end{equation}
where $\Sigma$ is the cost of a single evolution. 
Without loss of generality we will consider the cost of single gates, 
since similar arguments apply for the cost of $n$-qubit controlled gates. 
So, by performing superadiabatic probabilistic quantum computing, the average energetic cost is given by
\begin{equation}
\bar{\Sigma}_{SA}^{sing}\left( \omega \tau ,\theta _{0}\right)
=\left\langle N\right\rangle \Sigma _{SA}^{sing}\left( \omega \tau ,\theta
_{0}\right) =2\hbar \omega \csc ^{2}\left( \frac{\theta _{0}}{2}\right) 
\sqrt{1+\frac{\theta _{0}^{2}}{4\left( \omega \tau \right) ^{2}}} \text{ \ .}
\label{MediaCost}
\end{equation}%
The function $\bar{\Sigma}_{SA}^{sing}\left( \omega \tau ,\theta _{0}\right) \rightarrow \infty$ as $\theta_0 \rightarrow 0$ 
and exhibits a minimum in the interval $0 < \theta_0 < \pi$ as a function of $\omega\tau$. Indeed, the angle $\theta_0^{\min}$ that minimizes $\bar{\Sigma}_{SA}^{sing}\left( \omega \tau ,\theta _{0}\right)$ 
grows monotonically with $\omega\tau$, with $\theta_0^{\min} \rightarrow \pi$ as $\omega\tau \rightarrow \infty$ (adiabatic limit).
Then, optimizing $\bar{\Sigma}_{SA}^{sing}\left( \omega \tau ,\theta _{0}\right) $ for $\theta_0$, we obtain
\begin{equation}
\frac{\partial}{\partial\theta _{0}} \bar{\Sigma}_{SA}^{sing}\left( 
\omega \tau ,\theta _{0}\right)  =\eta \left( \theta _{0},\omega
\tau \right) \left\{ \theta _{0}-\left[ 4\left( \omega \tau \right)
^{2}+\theta _{0}^{2}\right] \cotan\frac{\theta _{0}}{2}\right\} = 0\text{ \ ,}
\label{derivada}
\end{equation}%
where we have defined the function
\begin{equation}
\eta \left( \theta _{0},\omega \tau \right) =\frac{\csc ^{2}\left(
\theta _{0}/2\right) }{2\left( \omega \tau \right) ^{2}\sqrt{1+\theta
_{0}^{2}/4\left( \omega \tau \right) ^{2}}}.
\end{equation}
Note that $\eta \left( \theta _{0},\omega \tau \right) $ is nonvanishing in
the whole interval $\mathcal{I}_{\theta _{0}}\in \left[ 0,\pi \right] $. Thus, to obtain
the critical angle $\theta_0^{\min}$ in $\bar{\Sigma}_{SA}^{sing}\left( \omega \tau ,\theta
_{0}\right)$, we use Eq.~(\ref{derivada}) to note that $\omega \tau $ satisfies%
\begin{equation}
\omega \tau =\frac{\sqrt{\theta _{0}^{\min }}}{2}\sqrt{\tan \left( \frac{%
\theta _{0}^{\min }}{2}\right) -\theta _{0}^{\min }} \text{ \ ,} \label{wtTheta}
\end{equation}%
where we can see a dependence of $\theta _{0}^{\min }$ on the choice of $\omega \tau$. 
In addition, note that $\theta _{0}^{\min }$ is such that 
$\tan \left( \frac{\theta _{0}^{\min }}{2}\right) \geq \theta _{0}^{\min }$,
since the quantity $\omega \tau $ is required to be real and positive. 
The probabilistic advantage is plotted in Fig.~\ref{fig1}, where it is shown that the optimal 
value for $\theta _{0}$ is a continuous 
function of $\omega\tau$, being distinct 
of the deterministic implementation $\theta_0=\pi$. 
In the inset, we show the global minimum of the average energy for $\omega \tau = 0.01$, which occurs 
for $\theta_0 < \pi$.
In particular, 
 $\theta_0^{\min}$ moves away from $\pi$ as $\omega\tau$ is lowered, i.e., in the strong 
superadiabatic regime. 
As $\omega\tau$ shifts towards the adiabatic limit, we find that 
$\theta_0^{\min} \rightarrow \pi$.  
The optimization of the energy cost  is shown in the lower inset, where we define the 
fraction of energy  required by the optimized probabilistic model as a function of $\omega \tau$ as
\begin{equation}
\Sigma _{\text{rel}}\left( \omega \tau \right) = \frac{\bar{\Sigma}%
_{SA}^{sing}(\omega \tau ,\theta_0 ^{\text{min}})}{\Sigma _{\text{SA}%
}^{sing}\left( \omega \tau ,\pi \right) }.
\label{sigma_rel}
\end{equation}
Notice that $\Sigma _{\text{rel}}\left( \omega \tau \right)$ decreases in the superadiabatic regime, implying into 
a large reduction of the energetic cost for small values of $\omega \tau$. On the other hand,  $\Sigma _{\text{rel}}\left( \omega \tau \right) \rightarrow 1$ 
in the adiabatic limit, since $\theta_0^{\min} \rightarrow \pi$.

\subsection{Superadiabatic quantum search}

Here, we derive a superadiabatic Hamiltonian $H_{SA}(s)$ for the oracular quantum search governed by the adiabatic Hamiltonian $H_0(s)$ in Eq.~(\ref{eq:grover-hamiltoniano}). 
We will adopt linear interpolation, with 
$f(s)=1-s$ and $g(s)=s$, as in Ref.~\cite{farhi1} and write 
$H_{SA}(s)=H_0(s)+H_{CD}(s)$. In order to determine the counter-diabatic Hamiltonian $H_{CD}(s)$, we observe 
that, since $H_0(s)$ has real eigenstates, we use that $\langle \dot{n}(s) | n(s) \rangle = 0$ in Eq.~(\ref{CDH}), which implies that 
\begin{equation}
H_{CD}(s) =\frac{i \hbar}{\tau} \sum_{\xi=\pm} |\dot{E}_\xi (s) \rangle \langle E_\xi (s) | ,
\end{equation}
where the energies $|E_\pm (s)\rangle$ are given by Eq.~(\ref{eq:eivec}) and 
\begin{equation}
| {\dot E_\pm(s)}\rangle = -\frac{(N-1)b_\pm \dot b_\pm}{(1+(N-1)b_\pm^2)^{3/2}}|{m}\rangle  +  \frac{\dot b_\pm}{(1+(N-1)b_\pm^2)^{3/2}}|{\phi}\rangle.
\end{equation}
Note that only the ground and first excited states contribute to $H_{CD}(s)$, since the higher energy 
degenerate sector $\{|E^k_{\text{deg}}\rangle \}$ is composed by time-independent eigenvectors 
[see Eq.~(\ref{E-deg-Grover})]. Note also that the counter-diabatic Hamiltonian will naturally be non-oracular [see Eq.~(\ref{NO-eq})], 
with contributions from operators such as $|\phi\rangle \langle m|$ and $|m\rangle \langle \phi |$. This is the reason behind 
the time complexity $O(1)$ for the superadiabatic Hamiltonian. Naturally, such a result leads to an artificial approach. 
In a more physical scenario, superadiabaticity could be applied to the quantum search via the direct implementation 
of the Grover quantum circuit, through the controlled evolution approach discussed in Section~\ref{CSEandUQC}. 

\subsection{Energy-time complementarity in the quantum search}

Let us now analyze the time-energy complementarity relationship in the adiabatic and superadiabatic versions of the 
Grover search. In the adiabatic regime, the energetic cost can be computed from Eq.~(\ref{cost.5}) and using 
$\tau \rightarrow \infty$. Therefore, the adiabatic cost can be written as
\begin{equation}
\Sigma _{\text{ad}} = \int_{0}^{1} ds \sqrt{\sum_{m}%
\left[ E_{m}^{2}\left( s\right) \right] }  = \int_{0}^{1} ds \sqrt{E_{+}(s)^2 + E_{-}(s)^2+ (N-2) E_{\text{deg}}(s)^2}  \text{ \ .}  \label{cost.Grover1}
\end{equation}%
Let us initially consider the oracular Hamiltonian $H_0(s)$ in Eq.~(\ref{eq:grover-hamiltoniano}), 
whose eigenvalues are given by Eqs.~(\ref{Ener-lowest-G}) and (\ref{Ener-deg-G}). 
By considering the case of local adiabatic evolution provided by the interpolation in Eq.~(\ref{LAE-Grover}) and 
by taking $N\gg 1$, we obtain $E_{\pm}(s)\sim E_{\text{deg}}(s) \sim O(1)$, which implies from Eq.~(\ref{cost.Grover1}) 
into an energetic cost $\Sigma _{\text{ad}}^{LA} $ that scales as $O(\sqrt{N})$.  
On the other hand, in the superenergetic version of the quantum search, we adopt the 
interpolation in Eqs.~(\ref{eq:f-def})~and~(\ref{eq:g-def}). Then, by taking $N\gg 1$, we obtain now 
$E_{\pm}(s) \sim E_{\text{deg}}(s) \sim O(\sqrt{N})$, which implies into  
an energetic cost $\Sigma _{\text{ad}}^{SE}$ that scales as $O(N)$. This higher energetic cost is a consequence 
of the complementarity between energy and time, which arises to compensate the constant time complexity $O(1)$ 
of the superenergetic version. Naturally, the composite energy-time complexity is kept constant for both cases. 
This overall complexity is reduced 
by taking non-oracular artificial Hamiltonians. In the case of the adiabatic NLNO model, we use the Hamiltonian 
in Eq.~(\ref{NOH}), whose ground state and first excited state energies are given now by Eq.~(\ref{NOH-epm}), 
with the higher energies kept as in Eq.~(\ref{Ener-deg-G}). Its energetic cost $\Sigma _{\text{ad}}^{NO}$ can also 
be computed from Eq.~(\ref{cost.Grover1}) 
by considering the interpolation in Eq.~(\ref{NOH-inter}) and by taking $N\gg 1$. Then, we obtain 
$E_{\pm}(s)\sim E_{\text{deg}}(s) \sim O(1)$, which yields 
$\Sigma _{\text{ad}}^{NO}$ scaling as $O(\sqrt{N})$. 

For the superadiabatic algorithm, Eq.~\eqref{cost.5} must be used. Without loss of generality, we
set energy units such that $\hbar / \tau=1$. We find that for $N \gg 1$, the value of $\mu_\pm (s)$  in Eq.~(\ref{cost.5}) evaluate to 
\begin{equation}
\mu_\pm(s) = \langle \dot E_\pm(s) | \dot E_\pm(s) \rangle   =   \frac{(N-1) \dot b_\pm(s)^2}{(1+(N-1)b_\pm(s)^2)^{2}} ,
\end{equation}
which in turn gives the superadiabatic search energetic cost $\Sigma _{\text{SA}}$ of order $O(\sqrt{N})$, 
just reproducing the scaling of the NLNO adiabatic search. Similar results can be obtained if one chooses the 
spectral norm in the energetic cost, up to a common scaling factor $D^{1/2} = \sqrt{N}$ 
related to the dimension of the Hilbert space. These results are summarized in Table~\ref{table1}.

\vspace{1cm}
\begin{table}[h]
	\centering 
    \begin{tabular}{|l||c|c||c|}
        \hline 
        ~                            & Energy Cost       & Energy Cost      & Time Cost       \\ 
        ~                           & (Frobenius Norm) & (Spectral Norm)   & ~                      \\ \hline
        Local adiabatic     & O($\sqrt{N}$)       & O(1)                  & O($\sqrt{N}$)  \\  \hline
        Superenergetic     & O(N)                    & O($\sqrt{N}$)    & O(1)                \\  \hline
        NLNO                   & O($\sqrt{N}$)       & O(1)                  & O(1)                \\  \hline
        Superadiabatic     & O($\sqrt{N}$)       & O(1)                  & O(1)                \\  \hline
    \end{tabular}
\end{table}
\vspace{-0.1cm}
\textbf{\refstepcounter{figure}\label{table1} Table~\ref{table1}}. Energy-time complexity for several versions of oracular and non-oracular Hamiltonians for the Grover quantum search.

\section{Discussion}

We have discussed the energetic cost of shortcuts to adiabaticity and their consequences in 
quantum information processing. Specifically, we considered both the superadiabatic universal 
gate model via CE and the superadiabatic analog quantum search. For the gate model, we have 
shown that, differently from the adiabatic scenario, superadiabatic probabilistic gate implementations 
are energetically favorable with respect to deterministic gate implementations. This implies 
that the additional energy resources required by superadiabatic evolutions can be minimized by a 
suitable probabilistic model. Indeed, probabilistic evolutions have recently been considered in similar 
applications for QC. In particular, they have been used to cancel errors in adiabatic processes~\cite{Kieferova:14} 
and as a technique to decompose unitary operations~\cite{Paetznick:14,Bocharov:15}.  Here, we have shown 
a new aspect of probabilistic QC, which corresponds to an advantage in the energy balance for superadiabatic 
dynamics while keeping its performance for a fixed evolution time.  For analog quantum search, 
we have shown that the superadiabatic approach induces a non-oracular counter-diabatic Hamiltonian, 
with energy-time complexity equivalent to non-oracular adiabatic implementations. This explicitly shows that 
the Grover optimality is robust against transitionless drivings, which is reflected by a fixed energy-time scaling 
of the Hamiltonian.  

Implications of probabilistic superadiabatic QC under decoherence is a further challenge of immediate interest. 
In a quantum open-systems scenario, there is a compromise between the time required by adiabaticity and the decoherence time of the quantum device. Therefore, a superadiabatic implementation may provide a direction to obtain an optimal running time for the quantum algorithm while keeping an inherent protection against decoherence. In this context, it is our interest to understand to what extent decoherence can affect the 
optimal angle $\theta_0^{\min}$, investigating in particular if it can be robust against classes of decohering 
processes. Concerning specifically the Grover search, it would be interesting to understand whether 
superadiabatic implementations are equivalent to arbitrary non-oracular adiabatic Hamiltonians, as suggested in 
our present discussion. Moreover, the behavior of correlations 
such as entanglement and the investigation of experimental proposals in the superadiabatic scenario are also 
topics under investigation. 

\section*{Conflict of Interest Statement}

The authors declare that the research was conducted in the absence of any commercial or financial relationships that could be construed as a potential conflict of interest.

\section*{Author Contributions} 
All the Authors equally contributed to the conception of the work, development of main the results, 
and writing of the manuscript.

\section*{Acknowledgments}
M. S. S. thanks Daniel Lidar for his hospitality at the University of Southern California. 
I. B. C., A. C. S., and M. S. S. acknowledge support from CNPq / Brazil and the 
Brazilian National Institute for Science and Technology of Quantum Information (INCT-IQ).

\end{document}